# Multidomain lithium niobate microcantilever for smart biosensor.

Igor Ostrovskii, Andriy Nadtochiy, Keith Hollis

*University of Mississippi, University, MS 38677*(iostrov@phy.olemiss.edu)

*Abstract*: Fabrication of the bimorph lithium niobate cantilevers involves a micromachining of a z-cut wafer to create a multidomain structure. After etching, the bimorph ferroelectric cantilevers are fabricated at the locations near interdomain walls. Cantilever vibrations are excited by rf voltage due to piezoelectric effect in a ferroelectric chip, and optical detection is employed to pickup the vibrations. The self-assembling layers of two different Silanes are chemically deposited on the micro cantilevers. The layers are detected independently by two methods including a low frequency shift of cantilever natural vibrations, and Raman spectral lines from the Silane layers. Possible applications include smart biosensors in a real time domain.

1. The fabrication and characterization of micro- and nano-scale cantilevers have been a subject of extensive research efforts [1,2]. Several materials, such as silicon and it's compound $Si_3N_4$[3,4], carbon nanotubes[5], gold[6], magnetostrictive metal glass[7] are used for cantilevers fabrication. Piezoelectric ZnO film[8] and PZT ceramic[9,10] are reported in the kinds of composite cantilevers, in which a piezoelectric effect was used to excite vibrations. We cannot refer to known publications that use ferroelectric properties in this type of vibrators or ferroelectric-silane system as a component of a cantilever biosensor. At the same time, ferroelectric microcantilever has an advantage of being easily excited by applying an rf voltage to it.

In this work, we report our experiments on fabrication micro cantilevers from widely used ferroelectric material $LiNbO_3$ (LN), which has useful chemical-physical surface properties that allow binding the silanes to its surface as a component of a smart biosensor. It is well known that −Z plane of LN having Li¯ ions on top is characterized by much faster chemical etching rate than + Z side[11]. We use this peculiarity for chemical fabrication of a bimorph ferroelectric cantilever by submerging a multidomain ferroelectric crystal into an acid. Fabricated cantilevers are shown schematically in Fig. 1a. We use 500-µm-thick optically polished z-cut wafers of LiNO₃ available from the marketplace. They were diced on rectangular samples with the dimensions of ~10 mm x 10 mm. A central area of ~5 x 5 mm² is engineered to fabricate the inversely poled domains by applying dc electric field of 23 kV/mm. Domain width and period are 0.5 and 1 mm, respectively. After that, one side of the sample was cut out along the line between poled and unpoled regions. Remaining part of the specimen with a length of 1-3 mm is mechanically polished to get a thickness of ~0.1-0.15 mm in the multidomain area. Then the sample is etched in $HNO_3$:HF (1:1) solution at 100°C while all crystal bulk between the domain walls is dissolved. The cantilever vibrators are formed exactly in the locations of interdomain walls between two inversely poled domains, these walls are shown by dashed lines in fig. 1a. Each single cantilever is the bimorph ferroelectric microcrystal with antiparallel orientation of polar z-axes in two cantilever sides divided by the domain wall. It is shown in fig. 1a by two arrows directed up and down. The microphotograph of fabricated cantilever is given in Fig. 1b. The dimensions of fabricated cantilevers are in the range of 5 to 50 μm thick, 30 to150 μm wide, and 450 to 2500 μm long. A remaining multidomain part of the initial ferroelectric wafer (1 in Fig. 1a) is used to drive the cantilevers with the help of rf voltage applied to upper metal electrode 2, since the domain walls vibrate if rf voltage is applied to multidomain array[12]. We fabricated tens of samples with the cantilever masses ranging 300 ng to 100 mg, and resonance frequencies of natural/fundamental vibrations of 83120 Hz to 1920 Hz. All of the specimens demonstrate generally similar vibrational properties.

2. To measure the resonance vibration frequency, an optical system is used as shown in Fig. 1a. The experimental setup consists of a HeNe laser focused on the cantilever. A laser beam is partly covered by the cantilever, so that an output signal of the photo detector is modulated with cantilever frequency. The data taken are transferred through digital oscilloscope to computer. All experiments are performed at atmospheric pressure and room temperature. The resonance frequency is determined as a point of maximum amplitude in a frequency response curve presented in Fig. 2. The measured quality factor Q of the cantilevers is in the range of 100-1500. In our experiments, more thin vibrators usually have lower Q factor, which can be explained by a stronger influence of an ambient air on lighter cantilevers.

A mass sensitivity is investigated by putting a point-load on a free end of the cantilevers as shown Fig. 1a. We use the paraffin beads of 6 - 60 μm in diameter having their masses in the range of 0.1-100 ng, respectively, and our experimental technique is similar to those described in[13]. Paraffin ball has respectively low density and so bigger volume than other materials like metal or sand. Typical examples of the amplitude-frequency dependencies are shown in Fig. 2a. When the vibrator is loaded, the curve demonstrates the low frequency shift, from initial plot 1 to plot 2. The data from three cantilevers loaded by different paraffin beads with masses (Δm) from 100 pg to 31 ng are presented in Fig. 2b. The plots 1, 2 and 3 proof a good linear relationship between the load mass Δm and corresponding frequency shift Δf. The highest mass sensitivity for our small-dimension cantilever operating at normal conditions was found to be S = (−Δf/Δm) = 0.5 Hz/pg. In reality, we can increase that sensitivity with thinner cantilevers, but than an influence of ambient air may affect the measurements.

3. Surface modification of LN wafer by 3-aminopropyltriethoxysilane[14] and Chlorotrimethylsilane[15] have been previously done[14,15]. In our experiments with bimorph micro-cantilevers, we use Chlorodimethyloctadecylsilane (CDOS), and Z-6920 silane that is a mixture of polysulfidosilane components. Fresh cantilevers after etching are washed with deionized water and then dried. After that, the cantilevers frequencies are measured, and the calibration process is performed by putting the paraffin beads on the cantilever tips. Immediately after calibration, the cantilevers are silanized by immersing a whole LN platform into a 0.1 M solution of CDOS or Z-6920-silane in an methylene chloride for 2 hours. After silanization, the cantilever platform is washed with methylene chloride to clean out any non-bounded silane compound, and then dried. Z-6920-silane was obtained from Dow Corning, and CDOS and methylene chloride were received from Sigma-Aldrich.

To detect a silane layer onto cantilever by a vibration frequency shift, we need to know a mass sensitivity $S_U$ for uniformly covered vibrator.

$$S_U = -\frac{\Delta f_U}{\Delta m_U} \tag{1}$$

Theoretically, one can represent $S_U$ in the integral form of equation (2).

$$S_U = -\frac{\Delta f_U}{\Delta m_U} = \frac{1}{L}\int_0^L \sum_n B_n x^n dx \tag{2}$$

Where L is a length of the cantilever, $B_n$ are coefficients that depend on effective spring constant of cantilever, and x is a distance from clamp point ($0 \leq x \leq L$). To find the coefficients $B_n$, we make the experiments with a small paraffin bead positioned at different x, the results are shown in Fig. 3. As it is expected, a change in frequency Δf decreases when a position of load is changed from maximum x = L to the clamp point at x = 0. From the experimental plot of Fig. 3, we find that it is enough to hold three terms $B_n$ in equation (2). The magnitudes of these coefficients are: $B_1$ = 0.107, $B_2$ = -0.518, $B_3$ = 0.981. After we know the integral in equation (2) and the length of a given sample, we can calculate unknown mass of a deposited layer onto the whole surfaces of the cantilever. Actually, a uniformly deposited layer with a total mass $\Delta m_U$ gives a smaller frequency shift $\Delta f_U$ than those $\Delta f_P$ resulting from the same point mass $\Delta m_P$ deposited at $x$ = L. Then, the sensitivities for point mass at $x$ = L ($S_P$) and for uniformly distributed mass ($S_U$) are not equal. For our bimorph ferroelectric cantilevers, we find the ratio

$$S_U = (f_0 / 4.5 \cdot \Delta m_P) = 0.22\, S_P, \qquad (3)$$

Where $f_0$ is initial natural frequency before a load mass is deposited. Expression (3) is in good agreement with the calculations for composite PZT-steel[10] and magnetostrictive[7] cantilevers, where this coefficient is found to be 0.236. If we take a uniformly distributed mass $\Delta m_U$ as a certain effective mass that produces the same frequency shift as the point load at $x$ = L, which means $\Delta f_U = \Delta f_P$, then from the equation (3) we find for ferroelectric cantilever a relationship between the point mass and uniformly distributed mass with the same frequency shifts.

$$\Delta m_U = 4.5 \cdot \Delta m_P \qquad (4)$$

After we have the mass of Silane film $m_S$ onto cantilever surfaces with known area, we can estimate a surface density of the molecule groups. For example, typical cantilever LNZC6 with length x width of 1600x100 $\mu^2$ yields $m_S$ the surface density of 6 ± 2 groups/nm$^2$. The measured density of our film is in good agreement with a close-packed monolayer data of 4.5-5 groups/sq.nm[16]. We note that in some our experiments the silane films may have a higher density than one monolayer. This could be explaining by 3-D surface-induced polycondensation[16].

4. Alternative experimental test for Silane presence on the cantilever samples is Raman spectroscopy. To take Raman spectra at room temperature, we use PTI MD 5020 spectrometer with Hamamatsu R928 PMT detector. The 632.8 nm radiation from a 10-mW He–Ne laser with the plasma line rejection filter is used as the excitation source. Data acquisition time is 200 s for each spectrum. A resulting spectrum is obtained after averaging over 100 runs. The Raman spectra in the 2400–3300 cm$^{-1}$ range are presented in Fig. 4. According to well-known source[17], the bands observed at 2849 could be assigned to the $CH_2$ symmetric stretch vibrations, 2883 and 2891 cm$^{-1}$ to the $CH_3$ symmetric stretch vibrations, and 2916 cm$^{-1}$ to the $CH_2$ anti-symmetric stretch vibrations. Since the main Raman lines of $CH_{2,3}$-group vibrations are detected, we have a very strong evidence of the surface modification by the silanes.

5. A final test is to attach the nano-gold particles to DOD Z-6920 silane. Colloidal gold with 10 nm gold particles of 0.005 % concentration was purchased from Sigma-Aldrich. Neat solution is diluted to concentration about 0.0015% in deionized water. After surface modification by DOD Z6920-silane, the cantilever platform is submerged into the diluted colloidal gold solution. Following 2-hour treatment, the cantilever platform is rinsed in deionized water and dried. Than we measure a frequency shift, it turns to be 2.7-3.4 Hz for different cantilevers. The shift is reliably detectable since we can measure frequency change with an accuracy of ±0.05 Hz. We estimate a mass of the gold particles bounded to cantilever surface as 4±1 ng, which corresponds to a surface concentration of $(5.1±1.2) \cdot 10^3$ nano-gold particles per $\mu^2$.

6. ***In conclusions***, a new type of cantilever sensor is proposed. It is a ferroelectric bimorph vibrator covered with a Silane layer. Nano-gold particles are reliably detected by the ferroelectric-silane composite cantilever at normal pressure and temperature. The lithium tantalite cantilevers need more time for chemical processing, since this material has lower etching rate. Such cantilevers may be used to detect certain biological objects in real time measurements.

Acknowledgements

This work in part is made possible due to research grant FRP-2007, UM. We are thankful Drs. Mack Breazeale and Nico Declercq for useful discussions.

*FIGURES:*

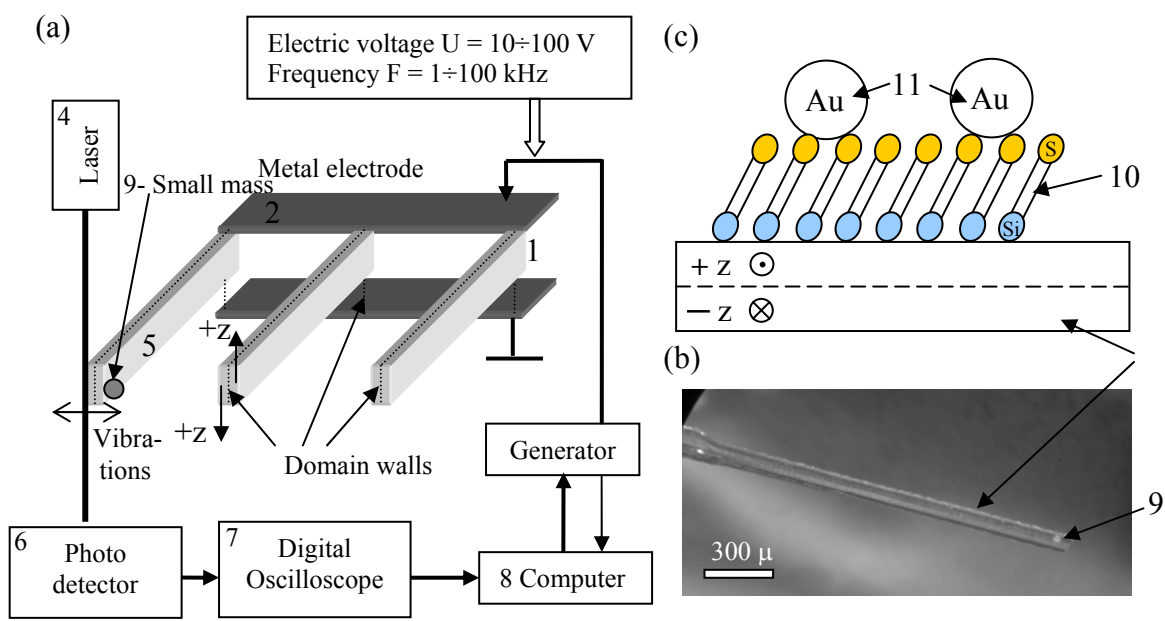

Fig. 1. (a) Schematic of the experimental setup. The micro cantilever platform (1) is clamped between two metal electrodes (2), and is driven by rf-voltage from a signal generator

(3). A laser light from (4) irradiates cantilever (5). Transmitted light is detected by photodiode (6), which transmits a signal into oscilloscope (7) followed by computer (8). (b) Microphotograph of cantilever (5) loaded with 28-μ-diameter paraffin bead (9). (c) Model of a bimorph cantilever with antiparallel polarization (±z) above and below an interdomain wall shown by dashed line. Single silane layer (10) modifies cantilever surface, and gold nanoparticles (11) are bounded by that layer.

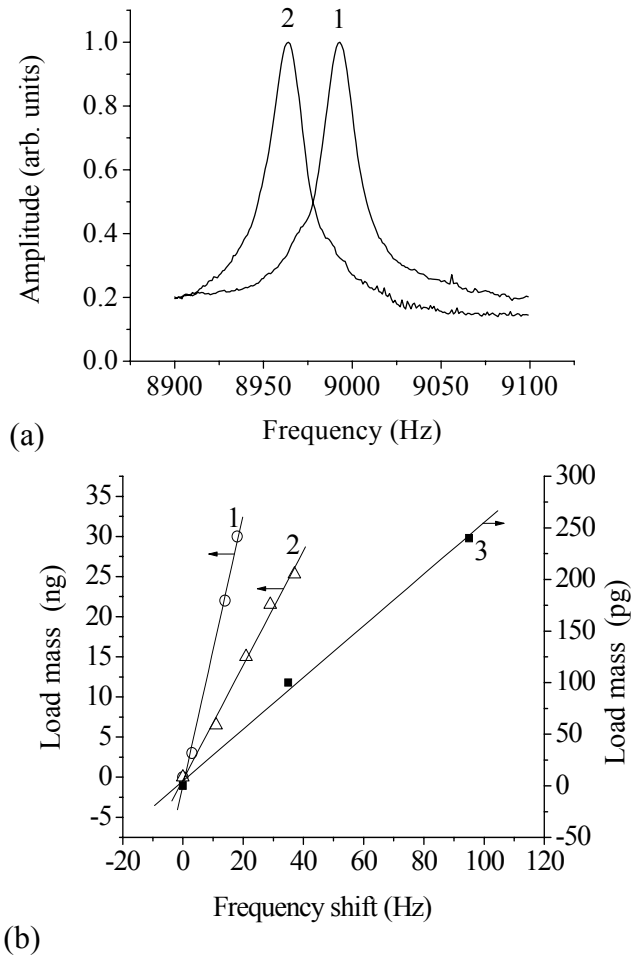

Fig. 2. Frequency dependencies and load mass calibration. (a) Vibration amplitude vs frequency from the sample LNZB-13. Plot 1 – free cantilever, in maximum F = 8993 Hz, 2 – loaded by 21.5 ng paraffin bead, in maximum F = 8964 Hz, that is 29 Hz low-frequency shift. (b) Load mass calibration. Plots 1 and 2 - nanogram-sensitive cantilevers LNZB-11 and LNZB-13, respectively; plot 3 - picogram-sensitive cantilever LNZA-3.

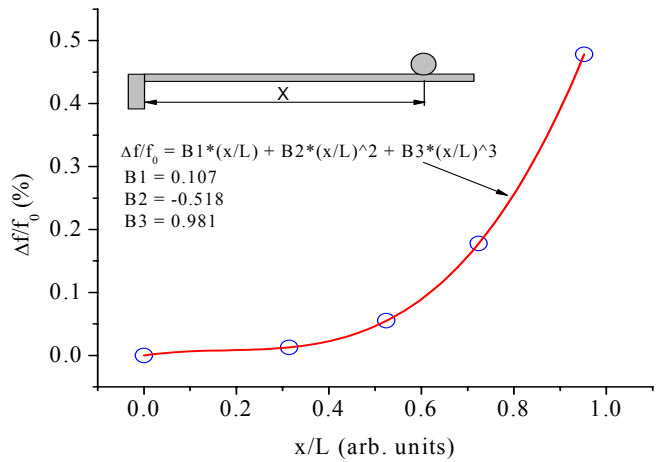

Fig. 3. Position sensitivity taken from the sample LNZB-14c1 loaded by 30 ng paraffin bead, which was put onto cantilever in 5 different points x/L.

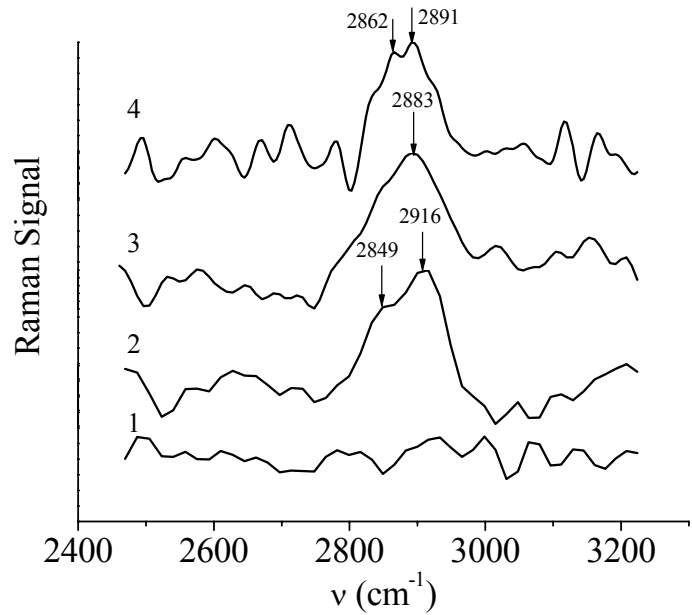

Fig. 4. Raman scattering spectra taken from various cantilever samples: Plot 1 – untreated sample LNZB28, 2 - LNZB28 after treatment by CDOS, 3 – LNZB23 modified with Z6920, 4 – LNZC6 modified with Z6920. Arrows show well-known Raman frequencies[11] of the $CH_2$ symmetric stretch vibrations at 2849-2861 $cm^{-1}$, $CH_3$ symmetric stretch vibrations at 2883-2884 $cm^{-1}$, and $CH_2$ anti-symmetric stretch vibrations at 2912-2916 $cm^{-1}$.